\documentclass[prx,twocolumn,superscriptaddress]{revtex4-2}
\usepackage{graphicx}
\usepackage{dcolumn}
\usepackage{bm}
\usepackage{xcolor}
\usepackage{soul}
\usepackage{amsfonts,footnote} 
\usepackage{amsmath}
\usepackage{physics,hyperref}
\usepackage{amssymb}
\usepackage[euler]{textgreek}

\begin{document}

\title{Chiral quantum optics: recent developments, and future directions}

\author{D. G. Sua\'rez-Forero}
\altaffiliation{Equally contributing authors}
\affiliation{Joint Quantum Institute, NIST/University of Maryland, College Park, Maryland 20742, USA}
\author{M. Jalali Mehrabad}
\altaffiliation{Equally contributing authors}
\affiliation{Joint Quantum Institute, NIST/University of Maryland, College Park, Maryland 20742, USA}
\author{C. Vega}
\affiliation{Instituto de Física Fundamental -
Consejo Superior de Investigaciones Científica (CSIC), Madrid, España}
\author{A. Gonz\'alez-Tudela}\email{a.gonzalez.tudela@csic.es}
\affiliation{Instituto de Física Fundamental -
Consejo Superior de Investigaciones Científica (CSIC), Madrid, España}
\author{M. Hafezi}\email{hafezi@umd.edu}
\affiliation{Joint Quantum Institute, NIST/University of Maryland, College Park, Maryland 20742, USA}

\begin{abstract}
Chiral quantum optics is a growing field of research where light-matter interactions become asymmetrically dependent on momentum and spin, offering novel control over photonic and electronic degrees of freedom. Recently, the platforms for investigating chiral light-matter interactions have expanded from laser-cooled atoms and quantum dots to various solid-state systems, such as microcavity polaritons and two-dimensional layered materials, integrated into photonic structures like waveguides, cavities, and ring resonators. In this perspective, we begin by establishing the foundation for understanding and engineering these chiral light-matter regimes. We review the cutting-edge platforms that have enabled their successful realization in recent years, focusing on solid-state platforms, and discuss the most relevant experimental challenges to fully harness their potential. Finally, we explore the vast opportunities these chiral light-matter interfaces present, particularly their ability to reveal exotic quantum many-body phenomena, such as chiral many-body superradiance and fractional quantum Hall physics.

\end{abstract}

\maketitle
\tableofcontents

\section{Introduction}

Generally speaking in physics, chirality refers to situations where the mirror image of a physical state does not possess the same properties as the original. This feature has many intriguing manifestations in the context of light-matter interactions, leading to various momentum- and spin-dependent phenomena in both photonic and electronic degrees of freedom. The definition of what is chiral varies across different fields. Therefore, in the next Section~\ref{sec:basic}, we begin by introducing the basic definitions and concepts that we will use later in this Perspective. The mechanisms section provides a concise introduction to the key concepts required to study chiral quantum optics. We then discuss recent developments in experimental platforms in Section~\ref{sec:plat}. In Section~\ref{sec:opportunities}, we present our view on potential future directions in the field of chiral quantum optics, ranging from single particles to many-body phenomena. We believe such a Perspective is timely, given the rapid evolution of this field since the visionary article by P. Lodahl \emph{et al}.~from 2017 \cite{lodahl2017chiral}. This evolution has occurred not only in experimental platforms but also conceptually, with new types of chiral light-matter interactions emerging beyond what was referenced.

\begin{figure}
    \centering
    \includegraphics[width=\columnwidth]{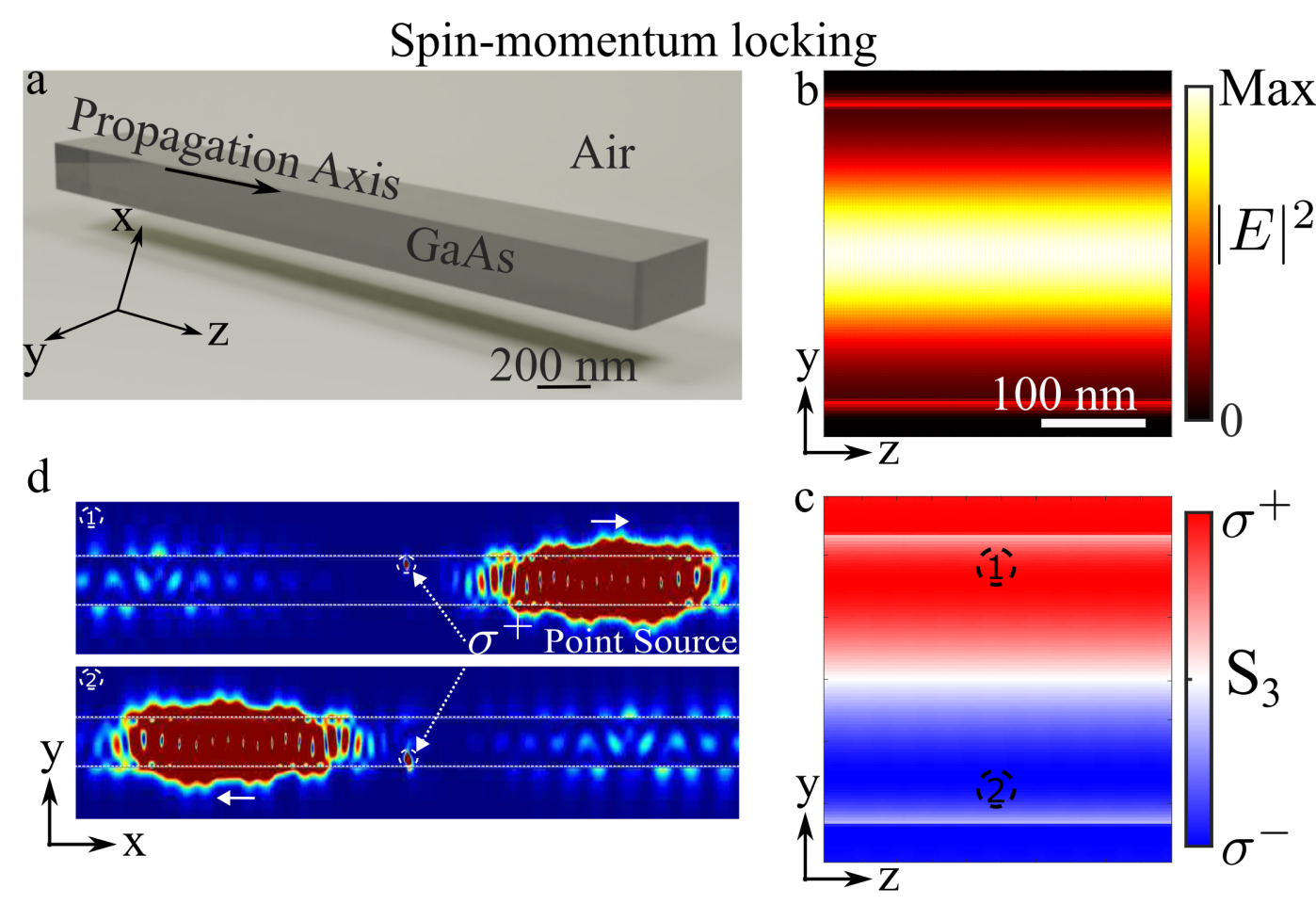}
    \caption{\textbf{Basic concepts in chiral quantum optics systems.} Upper panel: Spin-momentum locking. An electromagnetic mode propagating in the +z direction of a dielectric waveguide (panel a) has a distribution of electric field intensity shown in panel b. For concreteness, the waveguide's height and width are chosen to be  $140$ nm and $280$ nm, respectively, and the electric field propagation of wavelength $\lambda\!=\!940$ nm is considered. The polarization of such a mode is determined by a position-dependent superposition of its transversal and longitudinal field components with a relative phase that varies from $0$ to $2\pi$. This leads to regions where the direction of propagation and the circular degree of polarization are locked, as shown in panel c. $S_3\!=\!2\rm{Im}\{E_xE_{y}^*\}/|E|^2$ is the degree of circular polarization for the mode shown in b. In the presence of an external magnetic field that is perpendicular to the propagation direction, the circularly polarized transitions of the emitter are split. In this scenario, spin-momentum locking manifests as directional emission for each transition depending on the emitter's position. 
    (d) For a $\sigma^+$-polarized point source positioned in the upper (lower) part of the waveguide indicated by the dashed circle, the emission is preferentially in the right (left) direction of the waveguide. The emission direction will be reversed for $\sigma^-$-polarized point source (not shown).
    }    \label{sketch}
\end{figure}

\section{Basic concepts in chiral quantum optics~\label{sec:basic}}
For understanding and engineering chiral light-matter phenomena, several key concepts must be considered. These concepts describe symmetry considerations of internal and spatial properties of light and matter and their interaction. Here, we highlight two central elements: photonic spin-momentum locking and time-reversal symmetry breaking, as schematically depicted in the panels of Fig.~\ref{sketch}. Then, we also connect these two concepts with the emergence of non-reciprocity.

\subsection{ Photonic Spin-momentum locking}
According to Maxwell's equations, in the absence of charge, the divergence of the electric field is zero. This imposes constraints on the polarization of an optical field propagating in the $+z$ direction, as illustrated in Fig.~\ref{sketch}. These constraints manifest as photonic spin-momentum locking, i.e., a one-to-one relationship between transverse spin (polarization) and the direction of propagation. To be more explicit, let's consider a generic spatial profile for the electric field $\textbf{E}=\textbf{E}(x,y)\rm{exp}(-i k_z z)$. Applying the condition $\vec{\nabla}\cdot\textbf{E}=0$ requires the longitudinal electric field to be non-zero and equal to $E_z=\frac{-i}{k_z}(\partial_{x}E_x+\partial_{y}E_y)$. The non-vanishing $E_z$ component is out of phase with the transverse components $E_x$ and $E_y$ by $\pi/2$ (hence the factor $i$). Thus, the field can be elliptically or circularly polarized at certain locations. Such a spin-momentum locking is somehow analogous to helical edge states in the context of electronic spin quantum Hall physics, i.e. opposite spin components have opposite currents.

To illustrate this photonic concept, in Figure \ref{sketch}a, we show a typical dielectric waveguide structure made of Gallium Arsenide (GaAs). For a transverse-electric propagating mode with the intensity distribution displayed in panel b, there is a spatial dependence of the circular degree of polarization S$_3$ (panel c). Notice how the direction of propagation is locked to the light polarization in some regions. This implies that an emitter placed in those locations will have a directional emission depending on its spin (polarization) state. This effect is depicted in panel d, which shows the propagation of a pulse emitted from a $\sigma^+$-polarized point source. This is an example of a chiral optical decay of an emitter into a waveguide mode. The design and optimization of photonic devices capable of efficiently locking the direction of propagation to the circular degree of polarization is an active field of research. Some of these architectures will be discussed in the platforms section~\ref{sec:plat}.

\subsection{Time-reversal symmetry breaking}
Since Maxwell equations are invariant under time reversal, if $\Psi(t)$ is a solution, then its time-reversal partner $\Psi(-t)$ is also a solution. However, if time-reversal symmetry (TRS) is broken, either by an external magnetic or time-varying electric field, then time-reversal partner states cannot have the same property. For example, if an electric plane wave, with frequency $\omega$ and momentum $k$, expressed as $\textbf{E}(z,t)= \hat{x}e^{i\omega t +i k z }$, is a solution to the Maxwell equations, then it is not guaranteed that $\textbf{E}(z,t)=\hat{x} e^{-i\omega t +i k z }$ is a solution as well, at the same frequency. In other words, the opposite propagating solution might be absent or appear at a different frequency.
Note that such TRS breaking is local because the external field is held constant. Upon a global time reversal, the magnetic field would flip its direction.

We note that to obtain spin-momentum locking, the TRS is not required to be broken. For example, for a photonic waveguide, in the absence of an external magnetic field, both time reversal partners (opposite spins states propagating in opposite directions of the waveguide) have the same energy. In this scenario, the guided modes are referred to as \textit{helical} states. In contrast, in the presence of an external field and broken TRS, such states can be independently accessed and the system becomes \textit{chiral}. This can be achieved if the photonic material exhibits a magneto-optical response to the magnetic field or by coupling it to an emitter sensitive to external magnetic fields, such as through Zeeman splitting.

\subsection{Non-reciprocal master equation ~\label{subsec:nonreci}}

\begin{figure}
\includegraphics[width=0.4\textwidth]{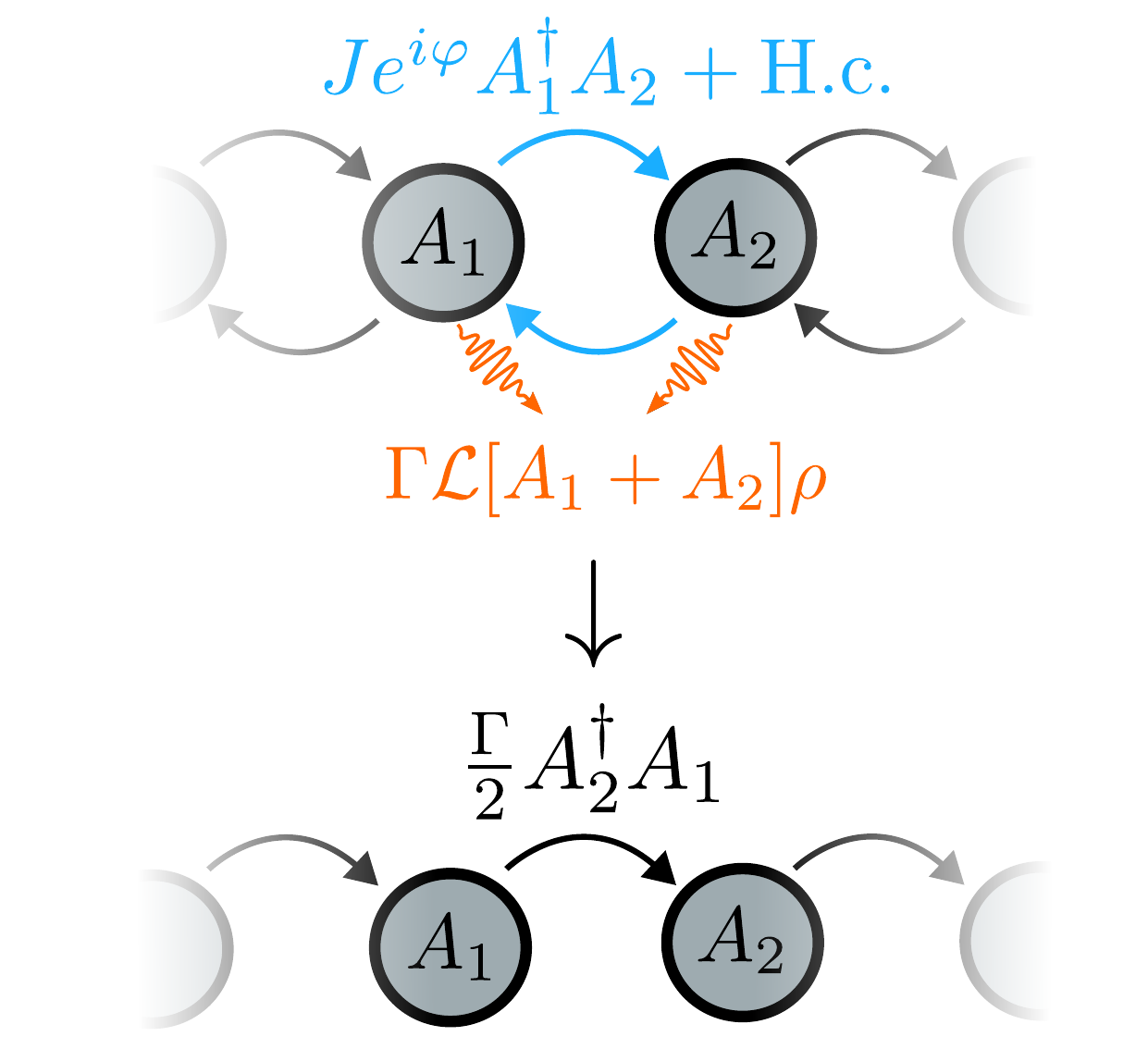}
\caption{\textbf{A toy model for achieving non-reciprocal interactions}. A pair of modes must simultaneously interact via both coherent tunneling (described by the Hamiltonian in blue) and collective decay (encoded in the dissipator in orange). If the tunneling (or jump operator) is complex, TRS is broken, leading to non-reciprocal dynamics.}
\label{fig:2cavities}

\end{figure}

Chirality and time-reversal symmetry breaking are also related to the emergence of non-reciprocal behavior. A simple model where this connection becomes evident is depicted in Fig.~\ref{fig:2cavities}.  It considers two bosonic modes, described by annihilation operators $A_{1,2}$, that coherently exchange excitations at a (possibly complex) rate $J$. The modes are also dissipatively coupled to a common bath with rate $\Gamma$ through a collective jump operator $A_1+A_2$. The equations of motion of the expected values, describing the field amplitudes in the two sites, are given by~\cite{metelmann2015a}:

\begin{align}
\frac{d\langle A_1\rangle}{dt} =& -\frac{\Gamma}{2}\langle A_1\rangle - \left(\frac{\Gamma}{2}+iJe^{i\varphi}\right)\langle A_2\rangle\,,\\
\frac{d\langle A_2\rangle}{dt} =& -\frac{\Gamma}{2}\langle A_2\rangle - \left(\frac{\Gamma}{2}+iJe^{-i\varphi}\right)\langle A_1\rangle\,.
\end{align}

We immediately observe that the system becomes non-reciprocal if the exchange couplings break TRS, i.e., $\varphi\neq 0$. Specifically, in the limiting case where $\varphi=\frac{\pi}{2}$ and $J=\frac{\Gamma}{2}$,  these equations reduce to:
\begin{align}
\frac{d\langle A_1\rangle}{dt} =& -\frac{\Gamma}{2}\langle A_1\rangle\\
\frac{d\langle A_2\rangle}{dt} =& -\frac{\Gamma}{2}\langle A_2\rangle - \Gamma \langle A_1\rangle\;,
\end{align}

which are manifestly non-reciprocal, since $\langle A_1\rangle$ affects $\langle A_2\rangle$, but not vice versa. The same situation can be obtained if the system is dissipatively coupled to a bath that breaks TRS, e.g., through a collective jump operator $A_1+e^{i\varphi}A_2$~\cite{metelmann2015a}. The two cases are connected via a gauge transformation $A_2\rightarrow e^{-i\varphi}A_2$.

This minimal model intuitively illustrates how non-reciprocity emerges from the interplay between TRS breaking coupling and collective dissipation to a common bath. Thus, the following Born-Markov master equation is the natural framework to describe this non-reciprocal system: 

\begin{align}
\frac{d\rho}{dt}&=-i\left[H_S+\sum_{i>j}\left(J_{i,j}\mathcal{O}_i^\dagger\mathcal{O}_{j}+\mathrm{h.c.}\right),\rho\right]\nonumber\\
&  +\sum_{i,j}\frac{\gamma_{i,j}}{2}\left(2\mathcal{O}_i\rho\mathcal{O}_j^\dagger-\mathcal{O}_i^\dagger\mathcal{O}_j\rho-\rho\mathcal{O}_i^\dagger\mathcal{O}_j\right)=\nonumber\\
&=-i\left(\rho H_\mathrm{eff}^\dagger-H_\mathrm{eff}\rho\right)+\mathcal{J}[\rho]\,,\label{eq:meq}
\end{align}

where $\rho$ is the emitter array density matrix, $H_S$ is the system's free Hamiltonian, $\mathcal{O}_i$ is the operator of the $i$-th local mode that couples to the bath, and $J_{i,j}$ and $\gamma_{i,j}$ are the effective coherent and incoherent photon-mediated interactions, respectively. In the last equation, we make a distinction between the effective non-hermitian evolution governed by $H_\mathrm{eff}=H_S+\sum_{i,j}(J_{i,j}-i\frac{\gamma_{i,j}}{2})\mathcal{O}_i^\dagger\mathcal{O}_j$ and the quantum jump terms $\mathcal{J}[\rho]=\sum_{i,j}\gamma_{i,j}\mathcal{O}_{i}\rho\mathcal{O}_j$. This formulation highlights the trace-preserving nature of the evolution, in contrast with purely non-Hermitian Hamiltonian formulations.

Notice that Eq.~\eqref{eq:meq} is valid for both reciprocal and non-reciprocal systems. The non-reciprocity exclusively depends on the coupling constants, i.e.~it takes place if $J_{i,j}\neq J_{j,i}$ and/or $\gamma_{i,j}\neq \gamma_{j,i}$. Such non-reciprocal couplings are present, for example, in chiral waveguide QED ~\cite{gardiner93a,carmichael93a,ramos14a,pichler15a,ramos16a,metelmann2015a,metelmann2015a,deBernardis2023} and chiral multi-mode systems ~\cite{Vega2023TopologicalQED,vega2024topological}. In both cases $J_{i,j}=\pm i\frac{\gamma_{1d}}{2}$ depending on whether $i\lessgtr j$, and $\gamma_{i,j}=\frac{\gamma_{1d}}{2}$. Interestingly, the coupling constants acquire an additional position dependence in the latter case. Other non-reciprocal mechanisms based on dissipative gauge symmetries have been recently proposed~\cite{wang2023a}. Independently of the nature of the non-reciprocity, these novel regimes open avenues to observe unprecedented quantum phenomena as discussed in~\ref{sec:opportunities}, in analogy to developments observed in classical many-body systems~\cite{Fruchart2021,avni2023}.

\section{Platforms for chiral light-matter interfaces~\label{sec:plat}}
Platforms that enable chiral light-matter interactions typically consist of a subset or all of the three following components: light, active materials, and photonic structures. For light, its spin angular momentum (SAM) and orbital angular momentum (OAM) are typically the degrees of freedom that play a central role in chiral light-matter interaction. In particular, $\sigma^+$ and $\sigma^-$ states which are the eigenstates of SAM, and vortex states of light which are the eigenstates of OAM are commonly used for inducing or manipulating chiral light-matter interaction. For photonic structures, devices that host helical counter-propagating guided modes are typically used to enable chiral light-matter interactions. This list includes 2D cavities, race-track, disk and PhC ring resonators, slab and photonic crystal (PhC) waveguides, and 0D open cavities. For active materials, systems with spectral, polarization, or OAM selection rules are most suitable for chiral light-matter interactions. Such systems include atoms and ions, solid-state and colloidal quantum dots (QDs), transition metal dichalcogenides (TMDs), and hybrid light-matter polaritons.

In this section, we review some of the emerging platforms where the integration of solid-state emitters in photonic structures enables chiral light-matter interactions. In particular, we focus on recent developments in the integration of III-V QDs, TMDs, and exciton-polaritons in photonic devices. These active materials have been interfaced with different photonic architectures to achieve chiral light-matter interaction. These implementations are summarized in Fig.~\ref{fig:review}, which compiles the different combinations of photonic structures and active materials where this has been achieved. The horizontal boxes categorize these demonstrations in 0D open cavities \cite{lyons2022giant,antoniadis2022chiral}, 1D waveguides \cite{coles2016chirality,Hallett2022,Barik2018,lodahl2017chiral,yang2019chiral,shreiner2022electrically,Klembt2018}, 
1D ring resonators \cite{lukin2023two,mehrabad2023chiral,ma2022chip}
and 2D cavities \cite{lobanov2015polarization,konishi2011circularly,suarezforero2023chiral,suarezforero2023}. The vertical boxes categorize each row based on the used active materials including QDs, TMDs, or polaritons integrated into each device. Specific details of each demonstration are discussed in the following sections.

\subsection{Photonic devices}

\subsubsection{2D Cavities}
This popular platform has been recently engineered to enable chiral light-matter interaction \cite{lobanov2015polarization,konishi2011circularly,suarezforero2023chiral,suarezforero2023}. In this system, a pair of parallel mirrors provide 2D confinement. Typically, they use distributed Bragg reflectors (DBRs) or metallic surfaces \cite{Megahd2022}; although a 2D cavity based on highly-reflective TMD monolayers has been recently reported \cite{suarezforero2023chiral}. In these architectures, external magnetic fields break the TRS for embedded QDs, TMDs, or QW excitons, leading to circularly polarized optical transitions. Therefore, light-matter interaction between the cavity modes and the embedded emitters becomes spin-selective. In other developments,  meta-surface-based chiral cavities \cite{semnani2020spin} and circularly polarized modes in liquid crystal cavities \cite{rechcinska2019engineering} have been demonstrated. They are designed to 
selectively reflect one circular polarization state of the incident light. Using these types of 2D chiral cavities, it was recently demonstrated that chiral scatterers can be detected via confined cavity modes \cite{bassler2024metasurface}.

Looking forward, several challenges need to be addressed in such 2D optical cavities for their widespread application. For example, TMD cavities need to achieve wider areas with chiral modes of high optical quality. Currently, they are still in the order of tens of $\mu$m$^2$. Moreover, strong magnetic fields are required to spectrally resolve the cavity modes due to the large linewidth of the TMD excitons that arise from inhomogeneous broadening. Linewidth reduction would therefore be advantageous for relaxing the required magnetic field strength. The development and improvement of fabrication techniques such as nano-squeegee \cite{rosenberger2018nano}, laser annealing, and chemical vapor deposition \cite{shree2019high,rogers2018laser} is systematically addressing these issues.

Regarding the metasurface-based chiral mirrors and cavities \cite{bassler2024metasurface,semnani2020spin}, Q factors are still modest and would benefit from enhancement, for example, towards achieving spin-selective strong-coupling physics recently reported in Ref.~\cite{suarezforero2023}. Another challenge has to do with emitter integration inside the cavity. Unlike chiral TMD cavities \cite{suarezforero2023chiral}, in which emitters can be readily embedded during the stacking of the TMD monolayers, metasurface-based chiral cavities might pose a challenge for similar emitter integration. However, recently exciting progress was reported for integrating chiral molecules in such cavities \cite{bassler2024metasurface}. 

\begin{figure*}
 \centering
 \includegraphics[width=0.99\textwidth]{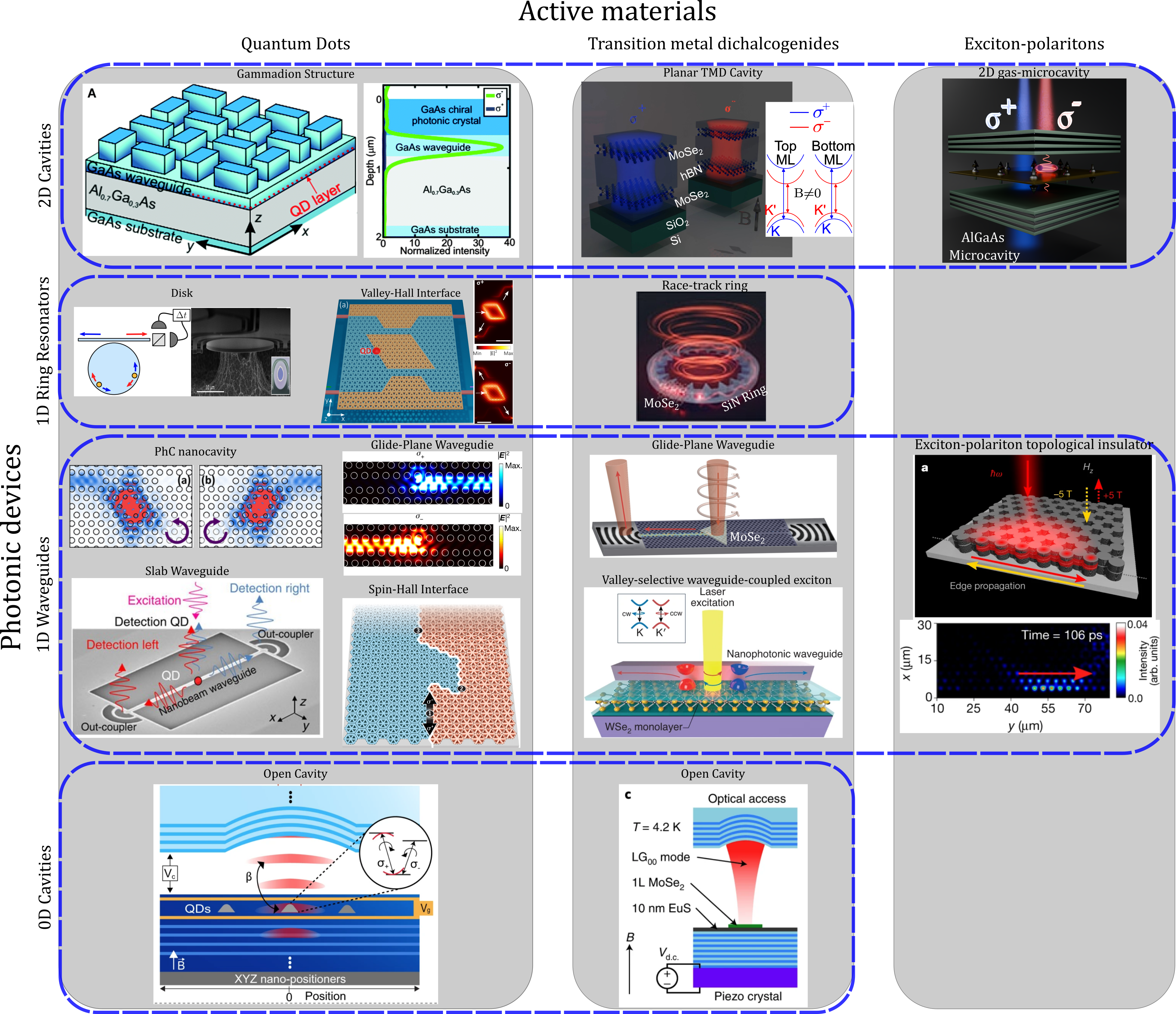}
 \caption{\textbf{Emerging platforms for chiral light-matter interaction.} Rows are associated with photonic architectures harnessed to induce this effect,
 while the active materials used for the matter component are compiled in the columns. Reprinted from Refs. \cite{lobanov2015polarization,lyons2022giant,antoniadis2022chiral,lukin2023two,mehrabad2023chiral,ma2022chip,coles2016chirality,Hallett2022,Barik2018,lodahl2017chiral,yang2019chiral,shreiner2022electrically,Klembt2018,lyons2022giant,antoniadis2022chiral}}
 \label{fig:review}
\end{figure*}

\subsubsection{Ring resonators}
Clockwise and counter-clockwise propagating modes in photonic ring resonators are another widely used chiral quantum optics platform \cite{lukin2023two,mehrabad2023chiral,ma2022chip,shomroni2014all,junge2013strong,scheucher2016quantum}. Examples include race-track nanobeam rings \cite{ma2022chip}, disks \cite{brooks2021integrated}, and spin- \cite{jalali2020semiconductor} and valley-Hall PhC ring resonators \cite{mehrabad2020chiral,barik2020chiral,mehrabad2023chiral,MJmehrabad2021}. In these systems, emitters are either embedded or evanescently coupled to the ring resonator's modes. To make the coupling chiral, the degeneracy of the optical transitions is typically lifted via an external magnetic field. This results in $\sigma^+$ and $\sigma^-$ emission from the emitters which will selectively couple to Counter-Clockwise or Clockwise modes of the ring resonators. Examples of such systems are shown in the second row of Fig. \ref{fig:review}. 

More recent efforts have been focused on further improving the efficiency of chiral coupling in such micro-resonators \cite{ma2024perfect}. The improvement arises from an engineered position-dependent backscattering of the whispering gallery modes. Another recent direction has been to increase the number of emitters coupled to the ring resonators, which was previously limited to one, in contrast with cold atoms systems, where large numbers of emitters have been chirally coupled to a photonic device \cite{liedl2023observation,johnson2019observation,pucher2022atomic}. In a recent exciting experiment, the coupling of two quantum emitters was reported for the realization of chiral collective decay physics \cite{lukin2023two}. Nevertheless, the true many-body chiral phenomenon with these systems lies still in its infancy. The main challenges include position-dependence chiral coupling efficiency of emitters into the resonator modes \cite{martin2024topological} as well as limited Purcell enhancement \cite{barik2020chiral}.

\subsubsection{Photonic waveguides}
One of the most popular platforms for chiral light-matter interaction is realized by integrating a two-level quantum emitter in a photonic waveguide that supports helical modes. Many types of optical waveguides have been developed in the last decades for chiral quantum optics. They include optical fibers \cite{petersen2014chiral}, nanobeam ridge waveguides \cite{coles2016chirality}, conventional line-defect photonic crystals (PhCs) \cite{coles2016chirality,sollner2015deterministic} and topological spin- \cite{barik2018topological,jalali2020semiconductor} and valley-Hall \cite{barik2020chiral,mehrabad2020chiral,martin2024topological,mehrabad2023chiral} PhCs. These waveguides typically support two counter-propagating optical modes, each for guiding light in one of the propagation directions of the waveguide. During the last few years, there has been intense research on integrating TMDs, QDs, and polaritons for the realization of 1D chiral light-matter interaction waveguides \cite{coles2016chirality,Hallett2022,Barik2018,lodahl2017chiral,yang2019chiral,shreiner2022electrically,Klembt2018}. Some of the key demonstrations of such devices for chiral quantum optics are shown in the third row of Fig.~\ref{fig:review}.

To address some of the challenges faced by 1D waveguides for chiral quantum optics, there are continued efforts to improve the efficiency of chiral light-matter interaction in these devices \cite{rosinski2024quantum,martin2024topological} as well as designing new chiral optics interfaces \cite{jalali2024strain}. Similar to ring resonators, the main challenge has to do with position-dependence of chiral coupling efficiency which hinders scaling up the system to a many-body regime. Moreover, regarding topological PhC waveguides, the non-trivial band structures of waveguide photonic crystals impose another challenge in designing structures with non-degenerate chiral modes. This is an important feature that provides robustness to the chiral modes, suppressing parasitic modes to which the light can scatter and be lost. Band structure optimization has become mandatory to address this issue, and important efforts are being made on this matter \cite{nussbaum2022optimizing,martin2024topological}.

\subsubsection{0D Open Cavities}

Another platform that enables chiral light-matter interaction is 0D optical nanocavities. Such devices include open cavities using Bragg mirrors \cite{lyons2022giant,antoniadis2022chiral}, in which a curved, mechanically-tuneable top mirror is used to confine light in all three spatial dimensions. Two examples of such devices are shown in the bottom row of Fig.~\ref{fig:review}. In these systems, chiral emission is generated from either TMDs or QDs integrated inside the open cavity.

Looking forward, it would be exciting to explore the implementation of the magnetic proximity effect to avoid the need for large magnetic fields~\cite{lyons2020interplay}. Quality factor improvement as well as better control over the cavity mode overlap with the emitters would also be advantageous towards this goal.

\subsection{Active materials}
So far we have discussed different photonic devices that enable chiral light-matter interactions. In this section, we provide a detailed discussion of the three main active materials used for chiral quantum optics.

\subsubsection{Quantum dots}
Having sub-diffraction limit sizes and atom-like selection rules, QDs are great candidates for chiral light-matter interactions. In particular, semiconductor QDs are attractive for their scalability, device integrability, and capability of delivering single photons with high purity and high repetition rates \cite{senellart2017high,dietrich2016gaas,kim2022development}. Motivated by these advantageous characteristics, chiral coupling of QDs has been demonstrated in many platforms such as glide-plane \cite{sollner2015deterministic} and W1 \cite{coles2016chirality} PhC waveguides. Moreover, nanobeam ridge waveguides have been utilized to demonstrate similar chiral coupling of QDs \cite{coles2016chirality}, as well as an optical spin readout of QDs' spin states \cite{javadi2018spin}. Chiral coupling of QDs has also been demonstrated in open cavities \cite{antoniadis2022chiral}. Recently, more exotic effects such as the direct observation of chirality-induced single-photon phase shift \cite{staunstrup2023direct} and the generation of entangled photon pairs from QDs' biexcitons embedded in a chiral waveguide \cite{ostfeldt2022demand} have been demonstrated.

Concurrently, topological photonic crystal waveguides \cite{mehrabad2023topological} also emerged as a novel spin-photon interface platform. These demonstrations include chiral coupling of solid state QDs to waveguides and ring resonators based on spin-Hall \cite{barik2018topological,jalali2020semiconductor} and valley-Hall \cite{barik2020chiral,mehrabad2020chiral,mehrabad2023chiral,martin2024topological,hallacy2024nonlinear} PhCs. Examples of key demonstrations of QD-integrated chiral quantum optics interfaces are shown in the left column of Fig.~\ref{fig:review}.

Applications for chiral light-matter interaction using QDs embedded in helical photonic structures are currently limited by the strong dependence of the interaction on the emitters' position. This dependence stems from the fact that the local circular polarization of guided modes in photonic waveguides has strong spatial variation \cite{hauff2022chiral,coles2016chirality}. It has been shown that even small deviations in the location of the emitter have severe implications for the chiral behavior of the system \cite{coles2016chirality,martin2024topological}. There are valuable efforts to address this issue using several techniques including site-controlled growth of semiconductor QDs, post-growth registration of their position, and inverse design approaches to find structures with good spin-momentum selectivity over wider areas \cite{martin2024topological}. The combination of the three mentioned approaches promises to unlock wider applications for chiral quantum optics.

Another challenge comes from the frequency-dependence of chiral coupling in photonic crystal waveguides \cite{lang2016time}. This dependency shows up in terms of the vanishing chiral coupling efficiency of emitters into the waveguide as one approaches their band edge. This issue is very disadvantageous since the favorable slow-light and the Purcell-enhanced emission effects occur near the band edge of typical W1 and Glide-Plane waveguides. In other words, one cannot have both high chiral coupling efficiency and Purcell enhancement at the same time, and the chiral coupling can only take place over a narrow bandwidth of the waveguides. Solving this issue is an active research field \cite{siampour2023observation}, but a final device capable of frequency-independent operation for chiral light-matter interaction remains elusive. Breaking the mirror symmetry in glide planar waveguides is currently the most common way to address this issue \cite{lodahl2015interfacing}.

The desired compactness of chiral devices imposes another challenge; finite-length waveguides present unavoidable stationary Fabry-Perot modes that deteriorate the chiral coupling of the emitters~\cite{coles2016chirality,ostfeldt2022demand}. Suppressing the undesired back reflections at the couplers of the waveguides becomes fundamental to reducing this effect. Coupling gratings with ultra-low back reflections have been demonstrated in semiconductor waveguides \cite{zhou2018high}, but a design with ultra-low back reflections, wide bandwidth, and high in-coupling and out-coupling efficiencies, remains an open problem. Other approaches aim to reduce the mode mismatch between the guided modes and the out-couplers, to minimize back reflections \cite{siampour2023observation}. Tapered interfaces offer additional capabilities such as the implementation of slow-light to fast-light structures. In this case, the chiral light-matter interaction is locally enhanced, and the generated light is then guided into fast modes, which are much more robust against Anderson localization from defects and disorder in the structure.

\subsubsection{Transition Metal Dichalcogenides}
The strong magnetic properties of excitons in transition metal dichalcogenides make these materials appealing for developing devices with chiral features. Demonstrations such as the spin-selective light-matter coupling of a TMD and a 0D cavity mode \cite{Lyons2022}, the fabrication of a chiral planar optical nanocavity made of TMD reflectors \cite{suarezforero2023chiral}, or the generation of quantum light with chiral features \cite{li2023proximity} were possible thanks to the large magnetic g factor of these materials. In these cases, the chirality comes from a TRS breaking induced by a strong magnetic field in the Faraday configuration (perpendicular to the sample plane) lifting the energy degeneracy between optical transitions of opposite spin. These demonstrations promise to unlock new ways to manipulate and study the light-matter interaction in a non-reciprocal platform.

The optical selection rules of these materials have been further harnessed by interfacing TMD emitters and helical photonic structures. Such demonstrations use the light emission from the excitonic recombination in the TMDs to show a preferential direction of propagation depending on its spin. There are different recent demonstrations of harnessing the chirality in both nanophotonic waveguides \cite{shreiner2022electrically} and topological photonic structures \cite{liu2020generation,li2021experimental,rong2023spin} to manipulate the direction of the light emission from a TMD based on its circular degree of polarization. Such integration of TMD materials and helical photonic structures recently allowed the demonstration of a monolayer WS$_2$ chiral laser at room temperature and zero magnetic field \cite{duan2023valley}. The central column of Fig.~\ref{fig:review} displays some of the mentioned examples of chiral light-matter interaction using TMD materials as active media.

Similar to the case of quantum QDs, the chiral light-matter coupling in these materials is sensitive to the location of the emitter inside the photonic structure. Given that the typically used photonic waveguides for TMD integration have waveguide modes spatially extending over large length scales (compared to the Bohr radius of the excitons), the chiral response is sensitive to the location of the excitation, for which more suitable helical photonic modes are required \cite{martin2024topological}. Moreover, the strength of the coupling between TMD excitons and photonic waveguides is typically low due to the weak evanescent electric fields of photonic waveguides, which encourages the search for better waveguide designs for on-chip TMD integration.
Other downsides of the platform are the inhomogeneous broadening, the typically small areas of the TMD monolayers with high optical quality, and the strong dependence of the excitonic resonances with the conditions of the embedding materials. Techniques such as the so-called "nano-squeegee" \cite{rosenberger2018nano} have greatly improved the quality of the interfaces in these devices, and advancements in the Chemical Vapor Deposition (CVD) growing technique allow the fabrication of larger monolayers with record-breaking optical qualities \cite{shree2019high,rogers2018laser}. Further, recent developments in sub-wavelength photonic structures that can host directional coupling offer more accurate future applications for chiral phenomena in these materials \cite{sarkar2024sub}. The fast evolution of the field allows one to envisage broad applications of chiral light-matter interaction using this platform as active media.

\subsubsection{Microcavity polaritons}
The magnetically-induced TRS breaking in exciton-polariton semiconductor structures has proven to be a powerful tool to create and engineer chiral physics in these systems. Their hybrid light-matter excitations are subject to the selection rules that lock the spin and polarization, which combined with a strong spin-orbit coupling has enabled the implementation of an exciton-polariton topological insulator \cite{nalitov2015polariton,klembt2018exciton} and a polaritonic system with spin-selective (chiral) light-matter coupling \cite{suarezforero2023}; both demonstrations are compiled in the right column of Fig.~\ref{fig:review}. The former demonstration uses honeycomb lattices of microcavity polaritons, while the latter one uses the unique advantages of the quantum Hall effect in a planar optical cavity, which provides topological features to a gas of charge embedded in a quantum well. Systems in this regime have been pushed further by enhancing the Coulomb interaction, which led to demonstrations in which an electromagnetic field interacts with integer and fractional correlated matter states of both electrons and holes \cite{ravets2018polaron,knuppel2019nonlinear,lupatini2020spin}. These demonstrations brought the physics of correlated electrons into the polariton physics field, opening exciting opportunities.  

In the aforementioned cases, the chiral features originate in the matter component of the polaritonic quasiparticles. However, the photonic counterpart can also endow the system with such properties via the optical spin hall effect, in which the polarization splitting of propagating modes acts as an effective magnetic field, rendering the modes helical \cite{leyder2007observation, kavokin2005optical}. 

The strength of the polariton platform resides in the hybrid nature of the hosted quasiparticles. While the photonic component provides high coherence, the excitonic counterpart makes these particles capable of entering strongly interacting regimes. This combination, endowed with chiral features, is expected to enable the ultrafast control of the optical non-linearities and the development of robust polaritonic circuits.

Broader applications of chiral quantum optics in polariton systems require the relaxation of two restricting conditions: the ultra-low operation temperatures and the high magnetic fields. For the first limitation, novel materials, with more robust properties against temperature-induced decoherence, are being explored \cite{dusel2021room,septembre2023design}. The requirement for large magnetic fields could be solved by engineering better optical properties for the devices. For example, magnetic substrates have been used recently to enhance the magnetic properties of a TMD material via proximity effect \cite{norden2019giant}.

Another challenge is related to the topological band gaps that can be achieved in polaritonic structures, which are still very small, rendering challenging any practical application. The refinement of fabrication techniques promises to unlock optical cavities with higher quality factors and larger light-matter modes overlap. A stronger photon-exciton coupling is mandatory for the development of more accessible chiral edge modes in this platform.

\section{From linear to many-body regime~\label{sec:opportunities}}

\begin{figure}
 \centering
 \includegraphics[width=0.99\columnwidth]{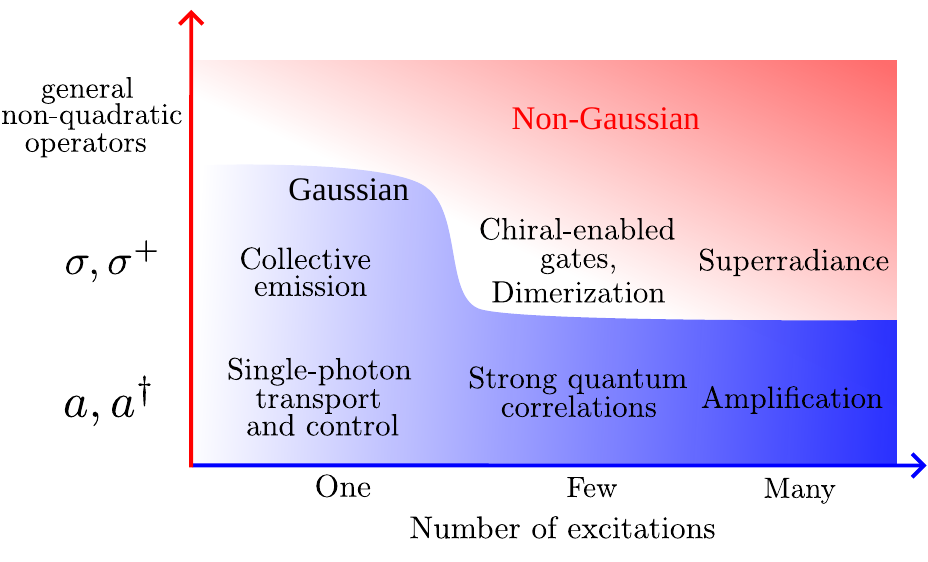}
 \caption{\textbf{Regimes of interest in chiral quantum optics}. Different recently explored phenomena in chiral quantum optics classified in terms of the operators involved and the number of excitations in the system. The vertical axis corresponds to the operators involved, namely if they are harmonic oscillator, ladder spin, or more generic forms of non-quadratic operators leading to non-Gaussian physics. The horizontal axis corresponds to the number of excitations, discerning between single-, few- and many-body phenomena.}
 \label{fig:regimes_of_interest}
\end{figure}

Irrespective of the particular realization, chiral quantum optical setups lead to non-reciprocal behavior as explained in Section~\ref{subsec:nonreci}. The interplay of such non-reciprocity with continuous or pulsed driving either through the photonic channel or the matter degrees of freedom opens opportunities to observe qualitatively new phenomena as we will review in this section. The complexity of such emergent phenomena depends on both the nature of the non-reciprocal modes, e.g., the operators appearing in the master equation \eqref{eq:meq}, and the number of excitations involved in them. These two factors are what we use in Fig.~\ref{fig:regimes_of_interest} to classify the complexity of these emergent phenomena. Currently, most experimental demonstrations lie within the lighter area of Fig.~\ref{fig:regimes_of_interest}. Their dynamics admit a Gaussian description in terms of single and two-body correlators~\cite{Walschaers2021}, either because they operate in regimes of low excitation numbers or because their associated operators are linear. This last criterion includes systems with large numbers of excitations but negligible quantum fluctuations, thus admitting a mean-field description. The exploration of the darker area of the diagram corresponds to the truly non-linear many-body regime, where the interplay between non-trivial operators, interactions, and many excitations makes Gaussian descriptions unsuitable. As we discuss the exploration of such a regime is still in its infancy with just a few theoretical and experimental works on the topic.

In what follows, we review both the experimental and theoretical exploration of these different regimes and classify them in terms of their prospective applications.

\subsection{Novel quantum light sources}

By the time the first review on chiral quantum optics was written~\cite{lodahl2017chiral}, it was already established that chiral light-matter interfaces lead to improvements in the control of light at the single-photon level (left-bottom corner of Fig.~\ref{fig:regimes_of_interest}). For example, employing emitters chirally coupled to one-dimensional waveguides opens the possibility to perform deterministic quantum state transfer protocols~\cite{Cirac1997} with considerable improvements~\cite{Yao2013, Dlaska2017, Lemonde2019, Mahmoodian2016,deBernardis2023}, to control the photon phase~\cite{sollner15a} and route~\cite{Shomroni2014,Ralph2015,Gonzalez-Ballestero2016}, to perform atom-photon swap operations~\cite{bechler2018passive}, and to induce non-reciprocal photon transmission~\cite{sayrin15a,scheucher2016quantum} and amplification~\cite{pucher2022atomic}, among other applications. However, more recently, the focus has evolved to investigate the implications of these chiral light-matter interfaces in multi-photon contexts (moving to the right in the bottom row of Fig.~\ref{fig:regimes_of_interest}). 

\begin{figure*}
 \centering
 \includegraphics[width=0.99\textwidth]{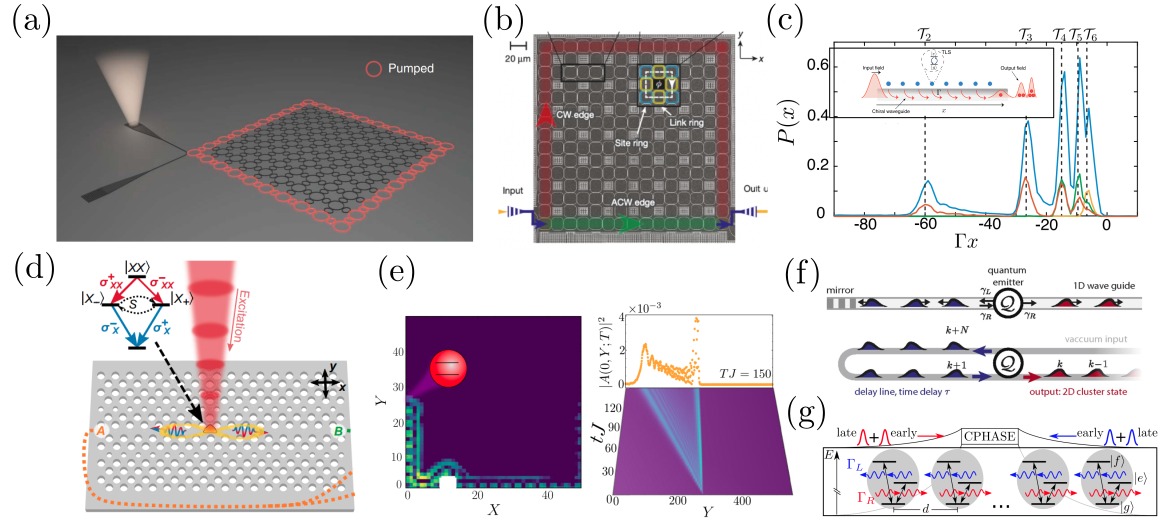}
 \caption{\textbf{Potential applications of chiral quantum light-matter interfaces to obtain novel photonic states and interactions.} Promising avenues include (a) topological lasing in a microresonator array~\cite{bandres18a} and (b) generation of topological photon pairs through spontaneous four-wave mixing~\cite{mittal2018topological}. Non-Gaussian correlations can be achieved with (c) multi-photon bound states in a platform with two-level systems chirally coupled to a photonic waveguide~\cite{mahmoodian2020dynamics}, (d) spontaneous emission of biexciton states~\cite{ostfeldt2022} or (e) single-photon states with time-bin entanglement between topological edge modes~\cite{Vega2023TopologicalQED}. (f) Cluster state generation using chiral waveguide coupling to $V$-level emitters~\cite{pichler17a} and sequential driving. (g)  Passive photonic phase gate enabled by chiral light-matter couplings to $V$-level emitters~\cite{schrinski2022passive}.}
 \label{fig:oppor}
\end{figure*}

Some of the first works going beyond the single-photon regime are those exploring the concept of \emph{topolasers}~\cite{Harari2018,kartashov19,Bandres2018,longhi2018,Secli2019,Amelio2020,zeng2020}, see Fig.\ref{fig:oppor}(a). These platforms combine an optically active medium with a system supporting chiral, propagating modes to achieve lasing. They were originally motivated by the robustness-to-disorder of topologically-protected chiral edge modes~\cite{Harari2018,Amelio2020} and the potential for lower operation thresholds~\cite{Harari2018}. However, these systems have shown new exciting features, such as a dependence of the relaxation time on the spatial extension of the gain towards steady-state~\cite{Secli2019} and the emergence of new amplification regimes~\cite{Secli2019}. One limitation of these systems is that in the lasing regime, the quantum correlations between photons are negligible, and hence their behavior is adequately captured by semi-classical theories.

Going beyond this semi-classical regime and inducing strong quantum correlations between photons are some of the most exciting prospects of chiral quantum optics nowadays. One way of inducing such correlations in these chiral channels is by directly exciting them with entangled photon light sources, e.g., using spontaneous parametric processes~\cite{mittal2018topological} (see Fig.~\ref{fig:oppor}(b)). However, this method limits the type of correlations that can be induced to that of the source. For example, in the pioneering experiments of Ref.~\cite{mittal2018topological} only Gaussian quantum correlations could be created. 

To overcome these limitations, one can harness the matter degrees of freedom. For example, two-level emitters are strongly interacting systems that cannot host more than one excitation at a time. This means that when several photons interact with chirally coupled emitters, the emitters can effectively induce strong interactions between the photons, resulting in non-Gaussian quantum correlations. This was predicted in recent theoretical works that studied the resulting transmitted photonic states in a system composed of two-level emitters chirally coupled to one-dimensional photonic waveguides~\cite{mahmoodian2020dynamics,olmos2020interaction}, see Fig.~\ref{fig:oppor}(c). The study showed how these systems support multi-photon bound states with varying propagation speeds depending on the photon number. Remarkably, this dependence results in the spontaneous separation of the different multi-photon components of a classical coherent state interacting with the ensemble. Other more recent works propose to exploit the emitter non-linearity and travelling-wave nature of the photons to design multi-mode beam-splitting operations~\cite{Lund2023} or deterministic multi-photon subtraction and addition~\cite{lund2024}.

An alternative method to generate non-trivial photon correlations by harnessing the matter degrees of freedom involves using emitters with richer energy levels than two-level systems. For instance, in a system where a quantum dot with biexciton states is chirally coupled to a one-dimensional waveguide, it was recently demonstrated that the spontaneous emission of biexciton states leads to the generation of polarization-encoded dual-rail photon pairs~\cite{ostfeldt2022}, see Fig.~\ref{fig:oppor}(d). These are examples of non-classical states of light with potential applications in quantum information protocols.

Finally, there are proposals exploring alternative mechanisms to create chiral quantum states of light without the need to increase the complexity of the matter's degrees of freedom. These methods include the use of sequential driving generation protocols~\cite{Gheri1998,Schon2005,Lindner2009,Economou2010,Schwartz2016}, multi-mode chiral systems~\cite{Skirlo2014,Skirlo2015,Vega2023TopologicalQED}, see Fig.~\ref{fig:oppor}(e)  or combinations of both, see Fig.~\ref{fig:oppor}(f). For example, Ref.\cite{pichler17a} theoretically demonstrates how a single $V$-level quantum emitter chirally manipulated using a mirror can be used to generate two-dimensional cluster states~\cite{pichler17a}, using the retardation delay as the second dimension. Conversely, in Ref.~\cite{Vega2023TopologicalQED}, coupling a single emitter to multiple edge states was shown to generate time-bin entangled states\cite{VanEnk2005}, see Fig.~\ref{fig:oppor}(e), whose entanglement complexity can be further enhanced using sequential driving protocols~\cite{Wein2022}.

\subsection{Chiral-enabled quantum gates}

Beyond the pure generation of non-trivial quantum states of light, chiral light-matter interfaces also enable engineering photon-photon gates in novel ways. For example, linear and non-linear optical operations were experimentally demonstrated in Ref.~\cite{Shomroni2014} and Refs.~\cite{volz2014,javadi2015} using chiral light-matter interactions. These ingredients can be combined with active and passive Gaussian optics to engineer deterministic controlled operations between photons using one~\cite{Ralph2015} or two chirally coupled emitters~\cite{Yang2022}. A way of avoiding these additional ingredients was put forward in a recent work~\cite{schrinski2022passive} which shows how an array of $V$-level emitters, in which the two optical transitions chirally couple the left and right propagating modes of a photonic waveguide, can introduce conditional phases between photonic qubits in a controllable manner without additional operations. In this theoretical proposal, they encode photonic qubits in time-bins (early/late), and then send two of them through the left or right side of the emitter array with opposite directions, see Fig.~\ref{fig:oppor}(g). In general, all the photonic modes, irrespective of the time-bin, will acquire a $e^{i\pi}=-1$ phase whenever they interact with an emitter, hence acquiring a global $(-1)^N$ phase after passing over $N$ emitters. However, if the early/late time bins of the left/right moving modes are chosen so that they overlap at a given emitter of the array, the overall state of the photons will acquire $\pi$ phase less since they cannot excite an emitter twice. This additional phase ultimately generates a passive controlled Z gate between the photonic qubits which, together with single qubit rotations, constitutes a universal resource for photonic quantum computation. An exciting perspective is to extend such control to chiral multi-mode waveguide setups, e.g., the ones appearing at the edges of topological photonic systems~\cite{Vega2023TopologicalQED}, where it has been recently shown one only needs two-level emitter non-linearities to get the controlled phase~\cite{LeviYeyati2024}, and where the photons would have the added value of topological protection for their propagation.

\subsection{Polariton quantum many-body phases}
One of the most exciting prospects of chiral light-matter interfaces is the possibility of discovering novel many-body phases of light that are absent in other systems. A paradigmatic example of this is the observation of a small-scale photonic Laughlin state~\cite{clark2020observation}, a long-sought but elusive goal in the photonic community. The key idea was to induce photonic interactions in a chiral cavity QED setup through coupling to a cloud of Rydberg atoms, as schematically depicted in Fig.~\ref{fig:oppor2}(a). However, stabilizing these states for large numbers of photons remains an outstanding challenge.

Beyond this milestone experimental observation, there have been other theoretical proposals for observing such many-body states. For instance, by strongly coupling emitters to two-dimensional Landau levels, one can find hybrid Landau-photon polaritons~\cite{DeBernardis2021} that inherit the chirality of the original Landau levels but feature a non-linear energy spectrum, as shown in Fig.~\ref{fig:oppor2}(b). 

Other exciting possibilities recently opened after demonstrations of optically-induced non-reciprocity in polariton condensates \cite{gnusov2023quantum,del2024non}. In this case, the challenge is to add strong particle correlations to the shown versatility of the optical control on the system's chirality.

Another intriguing future direction can be the realization of many-body chiral phenomena in a synthetic time multiplexed lattice. Recently, by engineering synthetic gauge fields and boundary edge states, an experimental method for the realization and manipulation of the directional flow of light in such synthetic space was developed \cite{lin2023manipulating}. Moreover, fast dynamic control over the flow of light has been also recently demonstrated in similar platforms \cite{zheng2024dynamic}. However, these experiments are in the linear regime and no photon-photon interaction has yet been reported for these systems. It would be exciting to explore the introduction of nonlinearity and its interplay with directional edge flows free from geometric constraints.

\subsection{Chiral many-body dissipative phenomena}

\begin{figure*}
 \centering
\includegraphics[width=0.99\textwidth]{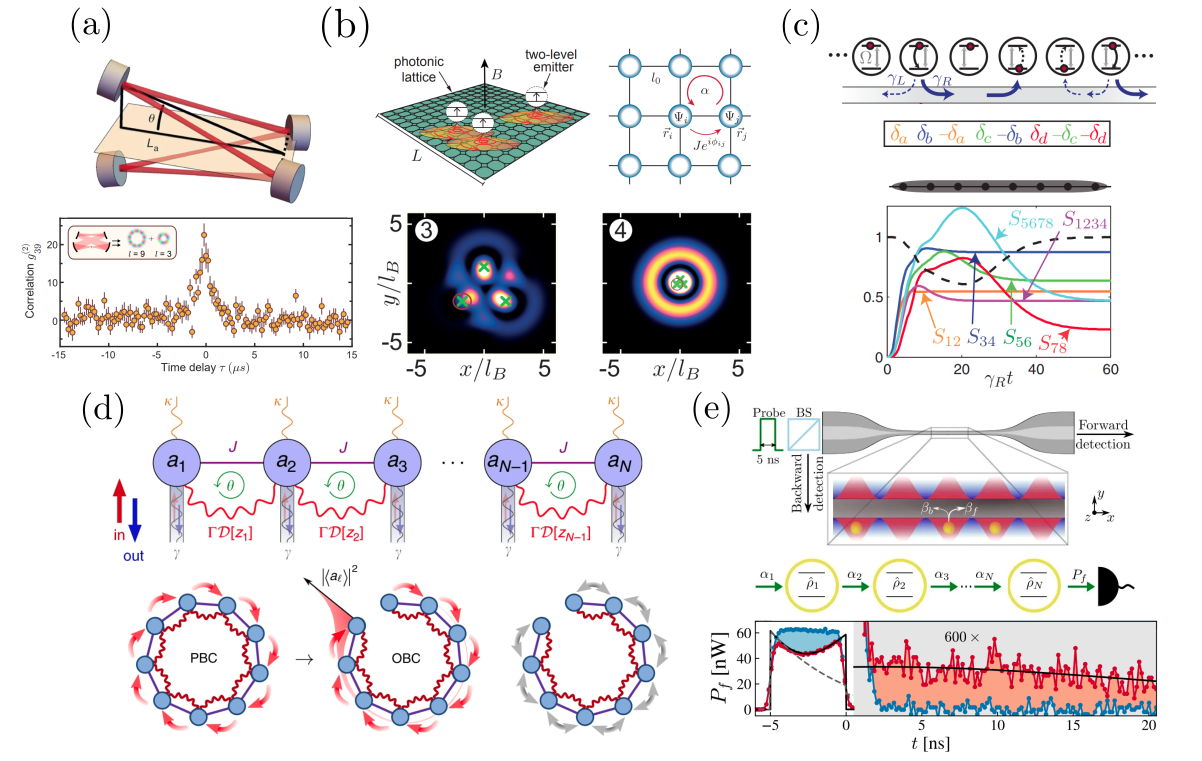}
 \caption{\textbf{Potential of chiral quantum light-matter interfaces to obtain novel many-body emitter phenomena, such as many-body or driven-dissipative phases.} (a) Laughlin states of optical photons through Rydberg atoms in a twisted optical resonator~\cite{clark2020observation, Schine2016}. (b) Landau-photon polaritons coupling quantum emitters to a two-dimensional photonic lattice~\cite{de2021light}. (c) Generation of pure multipartite entangled states in a chiral spin network~\cite{pichler15a}. (d) Directional amplification in driven-dissipative arrays of coupled cavities~\cite{Wanjura2020}. (e) Collective decay enhancement with atoms chirally coupled to a waveguide~\cite{Liedl2023CollectiveEnsembles}.}
 \label{fig:oppor2}
\end{figure*}

Beyond these purely coherent scenarios, another interesting direction is to consider the interplay of the non-reciprocal interactions of Eq.~\eqref{eq:meq} with continuous driving. This can result in qualitatively novel out-of-equilibrium dynamical processes and unconventional steady-states.  A first paradigmatic example of such potential applications is the prediction of the emergence of pure entangled many-body states in driven-dissipative scenarios~\cite{stannigel2012driven,ramos14a,pichler15a,ramos16a} (see Fig.~\ref{fig:oppor2}(c)). There, it was demonstrated how the interplay between chiral dissipation and a judiciously chosen driving for the emitters leads to a pure, multi-merized state, which can only emerge in such cascaded setups.

In another potential direction, non-reciprocal interactions can lead to non-trivial driven-dissipative photonic phases when they occur between photonic modes, for example, by chirally coupling cavities instead of emitters. Combining chiral photon tunneling and incoherent pumping (gain) or parametric driving was recently discovered to result in remarkable phenomena such as phase-dependent chiral transport~\cite{macdonald2018a,Karakaya2020,wonlnik2020} or topological amplification of steady-states~\cite{peano16a,Porras2019,wang2019a,Wanjura2020,Wanjura2021,Flynn2020,Flynn2921a,Ramos2021,GomezLeon2022a,GomezLeon2023,Ramos2022a, vega2024topological}. The latter cases are characterized by an exponential growth of the population with system size. This phenomenon can be understood from the emergence of topological edge modes in the dynamical matrix governing the system's dynamics, as illustrated in Fig.~\ref{fig:oppor2}(d). These non-Hermitian photonic lattices are particularly unique due to their sensitivity to the boundary conditions of the system, which can be exploited for quantum sensing~\cite{Lau2018,McDonald2020,budich2020,parto2023,sarkar2023}. They also become unstable in closed systems because of the continuous amplification of the population. However, interactions are expected to modify this ever-growing amplification behavior: an intriguing opportunity to observe novel dissipative quantum phase transitions, which remains a significant theoretical challenge. 

We conclude this section by mentioning that driving is not always required to obtain non-trivial out-of-equilibrium phenomena. Some initial states can already yield non-trivial dynamics, as in the atomic excitation trapping in dissimilar chirally-coupled atomic arrays \cite{handayana2023atomic} or the improvement of transient entanglement generation~\cite{Gonzalez-Ballestero2015}. However, the most interesting prospects lie in the many-body regime, which can be achieved, for example, by fully inverting the emitters. In such a case, one encounters a complex out-of-equilibrium situation due to the potential spontaneous build-up of correlations from photon-mediated interactions. In the non-chiral case, where photon-mediated interactions are purely dissipative, the fully inverted state was well-characterized by Dicke in the 1950s~\cite{dicke54a}, using the fact that all decay occurs through the symmetric subspace. In the chiral case, however, all symmetry subspaces become mixed, making it a numerically challenging problem. Recent experiments with atoms chirally coupled to optical fibers have demonstrated such collective enhancement~\cite{liedl2023collective}, as shown in Fig.~\ref{fig:oppor2}(e), along with superradiant bursts~\cite{Liedl2024} predicted before only for the standard Dicke model. The theoretical and experimental understanding of this chiral many-body regime remains an open question and is subject of on-going experimental~\cite{bach2024emergence} and theoretical research~\cite{windt2024effects,tebbenjohanns2024predicting}.

\subsection{Chiral light-matter interactions beyond dipole-approximation}
To explore the interplay between the chirality of electrons and the helicity of light, it was recently shown that OAM of light can be transferred to electrons in the quantum Hall regime \cite{session2023optical}. This observation can be understood as an analog to the optical pumping of bound electrons in atoms and molecules, in which the quantum states of atoms and ions are initialized and manipulated using a cyclical drive. In this recent experiment, optical transitions between electronic Landau levels are driven preferentially in an OAM-selective manner but are relaxed democratically. This demonstration opens an avenue to explore the interplay between light's vorticity and electron's chirality beyond the dipole approximation where the spatial profile of light becomes relevant for light-matter interaction. Beyond the quantum Hall regime, another recent demonstration showed the general experimental scheme for the generation of photocurrent in Graphene \cite{ji2020photocurrent} using OAM of light, which makes this research direction even more accessible.

It would be intriguing to expand these studies to other platforms such as GaAs. There, photoluminescence from excitons created by using light with OAM, may result in the generation of chiral photons. Another promising platform can be HgCdTe for which the groundwork has recently been laid  \cite{but2019suppressed}. In general, in semiconductors, electrons and holes bind because of coulomb attraction and OAM will be mainly transferred to the center of mass. This hinders the comprehension of the role of OAM \cite{grass2022two}. A complete understanding of such spatially coherent light-matter interactions requires further theoretical and experimental work.

\section{Conclusions and outlook }
In summary, the field of chiral quantum optics has experienced tremendous progress in the last five years. From an experimental point of view, many novel platforms have become available, opening the door to exploring new interaction regimes and their interplay with other quantum degrees of freedom, such as orbital angular momentum. Looking forward, one can foresee many more examples of such platforms becoming available, for instance, those based on subwavelength atomic or nanoparticle arrays \cite{peter2023chirality,kuznetsova2023engineering,peterchirality,Cerdan2023}. Moreover, similar ideas may be explored in other frequency regimes, e.g., microwave~\cite{owens2021chiral,rosen2024,Wang2024a}, with the potential to surpass the challenges of existing optical platforms. 

From the theoretical standpoint, many works have already demonstrated the potential of chiral quantum optics to facilitate the observation of exotic equilibrium and out-of-equilibrium many-body dynamics. However, there are still many outstanding questions that will trigger significant theoretical research and likely lead to new fundamental discoveries. A particularly intriguing direction is the generalization of these chiral light-matter regimes to photonic modes in higher dimensions, which some works are already exploring~\cite{Yu2020,GarciaElcano2023}, or their study beyond Markovian approximations~\cite{windt2024effects}. Connecting to this higher-dimensional direction, it is also intriguing to develop a topological classification of such non-reciprocal open quantum systems, going beyond purely quadratic (Gaussian) forms~\cite{villa2024a}. One of the ultimate challenges corresponds to the achievement of a truly non-linear, many-body regime for chiral systems (top, right corner of Fig.~\ref{fig:regimes_of_interest}), where unprecedented phases of light and matter are expected to emerge.

\begin{acknowledgments}
We gratefully acknowledge Edo Waks, Nicholas Martin, Arno Rauschenbeutel, Philipp Schneeweiss, and J\"urgen Volz for their insightful comments and discussions during the preparation of this manuscript. This work was supported by AFOSR FA9550-20-1-0223, FA9550-19-1-0399, FA9550-22-1-0339, ONR N00014-20-1-2325, NSF IMOD DMR-2019444, ARL W911NF1920181, Minta Martin and Simons Foundations. AGT and CV acknowledge support from the Proyecto Sin\'ergico CAM 2020 Y2020/TCS-6545 (NanoQuCo-CM), the CSIC Research Platform on Quantum Technologies PTI-001 and from Spanish projects PID2021-127968NB-I00 funded by MICIU/AEI/10.13039/501100011033/ and by
FEDER Una manera de hacer Europa, and TED2021-130552B-C22 funded by  MICIU/AEI /10.13039/501100011033 and by the European Union 
NextGenerationEU/ PRTR, respectively.
\end{acknowledgments}

\bibliography{bib.bib}
\end{document}